\documentclass[epsf,prb,twocolumn,showpacs,preprintnumbers]{revtex4}
\usepackage{graphics}
\usepackage{graphicx}
\usepackage{dcolumn} 
\usepackage{bm}
\usepackage{color}
\usepackage{epsfig}
\newcommand{\lnbn}{LaNiBN}
\newcommand{\lpbn}{LaPtBN}
\newcommand{\lmbn}{La${\cal M}$BN}

\begin{document}
\title{Electronic Structures and Phonon Spectra in Boronitride Superconductors
La{$\cal M$}BN ({$\cal M$}= Ni, Pt) 
}
\author{Myung-Chul Jung$^1$, Chang-Jong Kang$^2$, B. I. Min$^2$, and Kwan-Woo Lee$^{1,3}$}
\email{mckwan@korea.ac.kr} 
\affiliation{ 
$^1$Department of Applied Physics, Graduate School, Korea University, Sejong 339-700, Korea\\
$^2$Department of Physics, PCTP, Pohang University of Science and Technology, Pohang 790-784, Korea\\
$^3$Department of Display and Semiconductor Physics, Korea University, Sejong 339-700, Korea
}
\date{\today}
\begin{abstract}
We have investigated electronic structures and phonon spectra of newly discovered
isostructural superconductors LaNiBN ($T_c$ = 4.1 K) and LaPtBN ($T_c$ = 6.7 K).
We have found that their electronic structures are substantially three-dimensional, 
leading to metallicity both in NiB (PtB) and the intervening LaN layers.
Our {\it ab initio} phonon calculations show that almost all phonon modes contribute
to the electron-phonon coupling (EPC) mechanism, 
reflecting that both layers are involved in the superconductivity.
For \lnbn, we obtain an EPC strength of $\lambda$ = 0.52 and 
a logarithmically averaged characteristic phonon frequency of $\omega_{log}$ = 376 K, 
leading to $T_c$ = 3.9 K. 
Compared with the Ni $B_{1g}$ mode in \lnbn, the Pt $B_{1g}$ mode in \lpbn~ is reduced
by $\sim$70\%, leading to a slightly enhanced $\lambda$ = 0.56 and an $\sim$20\% reduced 
$\omega_{log}$.
The estimated $T_c$ is 5.4 K for \lpbn, in good agreement with the experiment.
We do not find any indication of magnetic instability for either \lnbn~or \lpbn,
which implies that both systems are EPC mediated superconductors.
Further, we have found an interesting trend of monotonic increase of $T_c$
with respect to the boron height in the NiB (PtB) layer 
of both borocarbide and boronitride superconductors, which suggests
a possible way to enhance $T_c$ in these systems.
\end{abstract}
\maketitle

\begin{figure}[tbp]
{\resizebox{6.5cm}{7.7cm}{\includegraphics{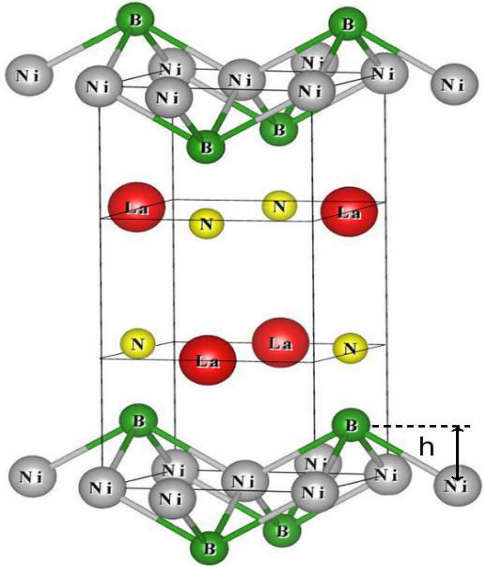}}}
\caption{(Color online) Crystal structure of La${\cal M}BN$,
which contains two formula units in the primitive cell. 
The inverse $\alpha$-PbO-type ${\cal M}$B layers are separated by two LaN layers.
In our relaxed structure, the Ni-B and Pt-B bond lengths are 2.17 and 2.30 \AA, 
respectively.
The ${\cal M}$-B-${\cal M}$ bond angles are 74.8$^\circ$ in \lnbn~
and 71.9$^\circ$ in \lpbn.
The symbol $h$ indicates the boron height in the ${\cal M}$-B layer.
 }
\label{str}
\end{figure}

\section{Introduction}
After the discovery of high-$T_c$ superconductors in cuprates and, more recently, in Fe pnictides,
layered systems containing transition metals (TMs), which in particular form a square lattice,
have been extensively studied.
These systems are of extreme interest due to their potential for being high-$T_c$ 
superconductors induced by the interplay between superconductivity and magnetic fluctuation.
In this point of view, boronitrides of (LaN)$_n$({$\cal M$}$_2$B$_2$) 
($n$ = number of layers, $\cal M$ = TM) and borocarbides of 
(${\cal R}$C)$_n$({$\cal M$}$_2$B$_2$) ($\cal R$ = rare-earth element), 
which exhibit various intriguing physical phenomena related to superconductivity 
and magnetism, have attracted immense attention.\cite{BC1,BC2}
In this research, we will focus on $n=2$ boronitride superconductors \lnbn~ and \lpbn, 
recently discovered by Imamura {\it et al.},\cite{hosono12} 
in order to understand the nature of their superconductivity. 

About 20 years ago, Cava {\it et al.} observed superconductivity 
in $n=3$ boronitride La$_3$Ni$_2$B$_2$N$_3$ at $T_c$ = 13 K.\cite{cava94-1} 
They synthesized $n=2$ boronitride \lnbn~too but failed to observe 
the superconductivity above 4.2 K.
\lnbn~showed ten times higher resistivity than the normal-state
La$_3$Ni$_2$B$_2$N$_3$.\cite{cava94-1}
Very recently, Imamura {\it et al.} successfully prepared a bulk superconducting 
\lnbn~sample with $T_c$ = 4.1 K, using a technique involving a high temperature of 1200 $^\circ$C and a high pressure of 5 GPa.\cite{hosono12}
Two isostructural superconducting compounds, \lpbn~and CaNiBN, were also synthesized with
$T_c$ = 6.7 and 2.2 K, respectively.\cite{hosono12}

These boronitrides show some distinctions as well as 
similarities in crystal structure compared to Fe pnictides. 
These systems commonly have inverse $\alpha$-PbO-type 
${\cal M}$-B layers, but in the $n=2$ boronitrides ${\cal M}$-B layers are separated
by two LaN layers with a width of 2.5 ($\pm0.1$) \AA, as displayed in Fig. \ref{str}.
As will be addressed below, boronitrides have metallic (LaN)$^0$ intervening layers,
while a superconducting Fe pnictide, LaFeAsO, possesses insulating (LaO)$^{1+}$ 
intervening layers. 

In the Ni-based borocarbides, which are isostructural with boronitrides, 
the interplay between superconductivity and magnetic fluctuation
has been extensively discussed.\cite{BC1,BC2}
Pt-based superconductors often show unconventional superconductivity
due to large spin-orbital coupling (SOC), incipient magnetism, 
and multigap nature.\cite{upt3,cept3si,LP05,li2pt3b,srpt3p,cho,anand}
Superconducting CePt$_3$Si and Li$_2$Pt$_3$B, which do not have inversion symmetry,
show significant SOC, possibly leading to the exotic superconductivity.\cite{cept3si,LP05,li2pt3b}
On the other hand, the effects of SOC in superconducting SrPt$_3$P are negligible.\cite{srpt3p}
So it is of interest to clarify whether the effects of SOC or magnetism matter
in \lnbn~and \lpbn.
However, except for the observation of superconductivity, other data in the literature 
are still very limited.

In this paper, we will address the electronic structures, including fermiology, which  
indicate no magnetic instability and negligible SOC effects in both superconducting boronitrides. 
Then, the phonon spectra and the electron-phonon coupling (EPC) will be discussed
to unveil the mechanism of the superconductivity.

\section{Structure and calculation method}
In the tetragonal structure of \lmbn(space group: $P4/nmm$, No. 129),
as shown in Fig. \ref{str}, 
the {$\cal M$} atoms lie on the $2b$ sites ($\frac{3}{4}$,$\frac{1}{4}$,$\frac{1}{2}$),
while the other atoms sit at the $2c$ sites ($\frac{3}{4}$,$\frac{3}{4}$,$z$).
Our calculations were carried out with the experimental lattice constants:
$a$ = 3.73 \AA~ and $c$ = 7.64 \AA~ for \lnbn~and 
$a$ = 3.8240 \AA~ and $c$ = 7.8649 \AA~ for \lpbn.\cite{hosono12,cava94-2} 
These values lead to an 8\% larger volume in \lpbn, 
consistent with the larger atomic radius of Pt.
The internal parameters $z$ were optimized using an accurate all-electron full-potential
code ({\sc fplo})\cite{fplo} 
because of the inaccuracy of the initial experimental data in \lnbn~ 
discussed earlier\cite{nat95} and the lack of available data for \lpbn.
The optimized values show tiny differences of at most 0.02 \AA~
between the local-density approximation (LDA) 
and the Perdew-Burke-Ernzerhof (PBE) generalized gradient approximation (GGA),\cite{pbe}
indicating that these values are nearly independent of the exchange-correlation functional.
In \lnbn, the optimized values $z$ are 0.1770 for La, 0.3545 for B, and 0.1636 for N.
The corresponding $z$ values of \lpbn~are 0.1666 for La, 0.3366 for B, and 0.1527 for N.
The differences in $z$ between the experimental and our relaxed values 
are substantial, 0.65 \AA~ for B and 0.13 \AA~ for N,\cite{cava94-2} 
but they are consistent with the relaxed values by Singh and Pickett 
for La$_3$Ni$_2$B$_2$N,\cite{wep95} 
showing differences of 0.45 \AA~ for B and 0.10 \AA~ for N.
Using our optimized values, the B-N bond lengths of 1.45 ($\pm0.01$) \AA~
are obtained for both systems, 
which are in agreement with the optimized values in the pseudopotenial code {\sc castep} 
by Imamura {\it et al.},\cite{hosono12}
whereas the ${\cal M}$-B bond length is 6\% longer in \lpbn.
These indicate that our optimized values are well converged.
In addition, by reducing the volume by 5\%, the B-N bond length remains unchanged,
implying strong B-N bonding and stiffness of the lattice (see below).
The structural parameters were optimized until forces were smaller than 1 meV/\AA.

For the full phonon calculations, we used the linear response method 
within the GGA implemented in the {\sc quantum espresso} package,\cite{qe}
which produces optimized internal parameters very close to those of {\sc fplo}. 
Our phonon calculations were carried out with the PBE ultrasoft pseudopotentials,
a Gaussian smearing factor of 0.05 Ry, a wave-function cutoff of 40 Ry, 
and a charge cutoff of 400 Ry. 
We used a $14\times 14\times 14$ $k$ mesh and a $2\times 2\times 2$ $q$ mesh 
for the phonon calculations
and a $28\times 28\times 28$ $k$ mesh for the EPC calculations, which require 
more careful treatment of the Fermi energy $E_F$.


\begin{figure}[tbp]
{\resizebox{8cm}{6cm}{\includegraphics{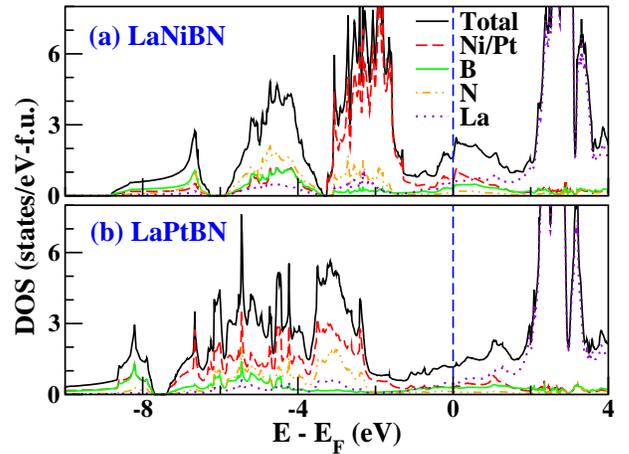}}}
\caption{(Color online) Total and atom-projected densities of states (DOSs)
of \lnbn~ and \lpbn.
\lpbn~ has 30\% smaller DOS $N(E_F)$ at the Fermi energy $E_F$
than \lnbn. $N(E_F)$'s are decomposed into roughly 40\% Ni, 30\% La, 20\% B, 
and 10\% N for \lnbn and into approximately 26\% Pt, 40\% La, 24\% B, and 10\% N
for \lpbn. Both systems have a van Hove singularity at $\sim$80 meV,
suggesting to enhance $T_c$ by small amount of electron doping.  
}
\label{dos}
\end{figure}

\begin{figure}[tbp]
{\resizebox{8cm}{6cm}{\includegraphics{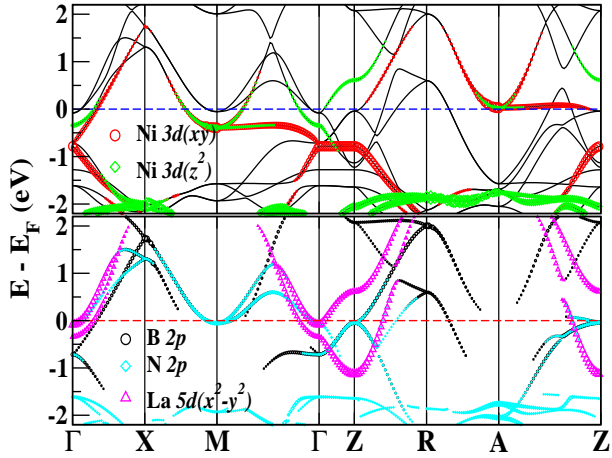}}}
\caption{(Color online) (top) Nonmagnetic band structure with Ni $3d_{xy}$ and $3d_{z^2}$
 fat bands of \lnbn~ in the optimized structure.
(bottom) Fat bands of B $2p$, N $2p$, and La $5d_{x^2-y^2}$ orbitals,
clearly indicating substantial B-N hybridization and a metallic LaN layer.
The horizontal dashed lines denote $E_F$.
The symmetry points are $\Gamma(Z)$ = (0,0,$\xi$), $X(R)$ = (1,0,$\xi$), and $M(A)$ = (1,1,$\xi$) 
in units of ($a/\pi$,$a/\pi$,$c/\pi$). Here, $\xi$ is 0 for the first symbols 
and 1 for the symbols in parentheses.
}
\label{ni_band}
\end{figure}

\begin{figure}[tbp]
\vskip 8mm
{\resizebox{8cm}{6cm}{\includegraphics{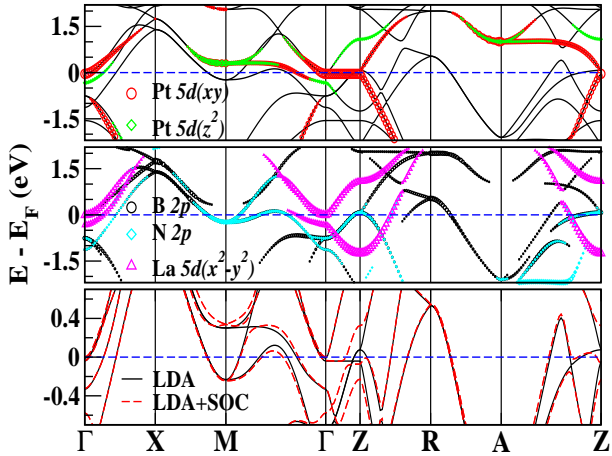}}}
\caption{(Color online) (top) Nonmagnetic band structure with Pt $5d_{xy}$ and $5d_{z^2}$
fat bands of \lpbn~ in the optimized structure.
(middle) Fat bands of B $2p$, N $2p$, and La $5d_{x^2-y^2}$ orbitals.
(bottom) Overlapped LDA and LDA + SOC band structures in the range of --0.7 to 0.87 eV, 
showing some distinctions along the $\Gamma$-$Z$ line just below $E_F$ and at the $M$ point
above $E_F$ but leading to negligible effects on the superconductivity (see text).
}
\label{pt_band}
\end{figure}

\section{Electronic Structures}
\subsection{Magnetic tendencies}
First, we address whether a magnetic instability is feasible 
in these boronitrides.
Fig. \ref{dos} shows the total and atom-projected densities of states (DOSs) for both systems.
DOSs at $E_F$, $N(E_F)$, are seen to be rather low, 1.65 and 1.17
(in units of states per eV per formula unit), which is
consistent with the bad metallic character observed in their normal states.
These values lead to bare specific-heat coefficients of 3.9 and 2.8 mJ/mol-K$^2$ 
for \lnbn~ and \lpbn, respectively, which are approximately a quarter of that of Fe pnictides 
in the nonmagnetic state.
In contrast to the Fe pnictides, this implies less magnetic instability,
or lack of magnetic instability, in these boronitrides.
Additionally, the contribution of Ni to $N(E_F)$ is small,  
0.33 states/eV-spin-f.u. for \lnbn. The contribution of Pt in \lpbn~ 
is only half of this value.
Using a typical intra-atomic exchange parameter of Ni $I\approx$ 1 eV,\cite{gunn} 
the Stoner factor $IN(E_F)$ is estimated to be 0.33 for \lnbn,
which is much smaller than unity.
So neither metamagnetism nor a near-stable ferromagnetic state is indicated. 
This feature is confirmed also by our fixed spin moment calculations 
in both systems (not shown here).
Again, no stable (anti)ferromagnetic states can be obtained from our calculations of 
both the local spin-density approximation and GGA for both systems.

Interestingly, considering a partial contribution of each atom at $E_F$,
$N(E_F)$ of La$_3$Ni$_2$B$_2$N$_3$ can be exactly reproduced from that of \lnbn.
This may imply a common mechanism of the superconductivity in both systems.
Note that La$_3$Ni$_2$B$_2$N$_3$ has higher $T_c$ by a factor of 3 and
larger DOS at $E_F$ in the unit cell by a factor of $\frac{3}{2}$ 
than \lnbn.\cite{wep95}

\subsection{Nonmagnetic electronic structures}
Now, we will focus on the nonmagnetic states, which represent the superconducting state.
For \lnbn, as shown in the DOS in Fig. \ref{dos}(a), 
Ni $3d$ states mainly lie in the range of --3.5 to --1 eV.
Below that regime, two separated manifolds have nearly equally mixed characters of B, N, and Ni ions
but relatively small La character.
Near $E_F$, the characters of all atoms appear, indicating metallicity in both NiB and LaN layers, 
as observed in La$_3$Ni$_2$B$_2$N$_3$.\cite{wep95} 
The Fermi energy pinpoints the midway of two small peaks, resulting from
the dispersionless bands with the mixed Ni $3d_{xy}$ and $3d_{z^2}$ characters
along the $M$-$\Gamma$ line at --0.5 eV and along the $A$--$Z$ line at 50 meV,
as displayed in Fig. \ref{ni_band}.
In many regimes near $E_F$, a mixture of these partially filled orbitals appears,
reflecting strong hybridization between inter- and intra layers.
The bottom panel of Fig. \ref{ni_band} obviously indicates the metallic La $5d_{x^2-y^2}$ band.

The total and atom-projected DOSs of the $5d$ counterpart \lpbn~ are given in Fig. \ref{dos}(b). 
As expected, the main part of Pt $5d$ orbitals is more widely spread
over the range of --7.5 to --2 eV, but some tail extends to 2 eV.
The total DOS smoothly increases near $E_F$, except for a van Hove singularity at 80 meV,
suggesting less sensitivity to carrier doping or moderate pressure.
The corresponding band structure and fat bands of the partially filled orbitals
are plotted together in Fig. \ref{pt_band}.
A comparison with those of \lnbn~ shows some distinctions in several regimes, in particular, 
around the $M$ point and along the $\Gamma$-$Z$ and the $A$-$Z$ lines.
At the $M$ point degenerate bands of B and N $2p$ orbitals move below $E_F$.
To compensate for this, the two fold degenerate flat bands 
with Pt $5d_{xy}$ and $5d_{z^2}$ orbital character shift up, 
being unfilled at the $M$ and $A$ points.
These bands are reflected as small peaks at 0.3 and 1 eV in the DOS in Fig. \ref{dos}(b). 
A more pronounced feature is the flat Pt $5d_{xy}$ band lying at --40 meV
along the $\Gamma$-$Z$ line, while the corresponding Ni band exists at --1 eV in \lnbn.
This leads to some differences in the phonon spectra, which will be addressed below.

These distinctions are also analyzed through interpretation of the occupation matrix,
which is useful for understanding the oxidation state.\cite{wep12}
In \lnbn, each orbital occupancy of Ni $3d$ is almost identical to 0.84$e$ per spin, 
resulting in a total of 8.42$e$. B $2p$ orbitals have 1.32$e$ in total.
As observed in the fat bands in Fig. \ref{pt_band}, 
in \lpbn~ Pt $5d_{xy}$ and $5d_{z^2}$ orbitals are
less filled by $\sim$0.04$e$, while the occupancy of Pt $5d_{x^2-y^2}$ increases a little.
Compared with \lnbn, 0.2$e$ transfers from the Pt ion to the B $2p$ ion,
whereas the occupancies of N $2p$ and La $5d$ remain unchanged.
So the Pt ions and the B ions are closer to  nominal values of Pt$^{2+}$ and B$^{2-}$
than the Ni and B ions in \lnbn, indicating less covalent bonding between the Pt and B ions.
(Of course, the formal charge concept is murky due to the metallicity and strong hybridization
in these systems.)
This may be consistent with the longer Pt-B bond length.
According to our calculations, however, the occupancies of \lpbn~ are nearly unchanged 
even for the same structure as in \lnbn, 
indicating that these distinctions are mainly due to chemical differences 
between Pt and Ni ions rather than differences in structure.

Since $5d$ systems often show substantial SOC effects, 
LDA + SOC calculations were performed for \lpbn.
As displayed in the bottom panel of Fig. \ref{pt_band}, however, 
a comparison of the band structures between LDA and LDA + SOC shows
negligible SOC effects near $E_F$, except for a small difference at the $Z$ point.
This is due to the wide width of the Pt $5d$ band.
Additionally, our calculations substantiated 
that the EPC strength and ,consequently, $T_c$ remain nearly unchanged, when including SOC.
As a result, the effects of SOC on the superconductivity of \lpbn~ are insignificant.
Now, we will exclude the SOC effects.  

\begin{figure}[tbp]
{\resizebox{8cm}{5.5cm}{\includegraphics{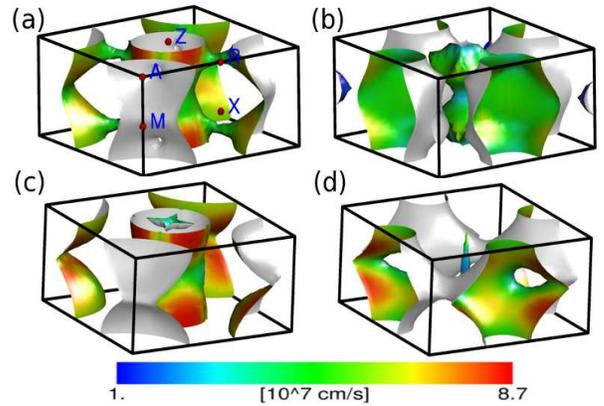}}}
\caption{(Color online) Fermi surfaces (FSs) of (a) and (b) \lnbn~ and (c) and (d) \lpbn~ 
in the optimized structure.
(For better visualization, FSs of each system are decomposed into two separate plots 
to reduce the hidden regime.)
In both systems, the $\Gamma$- and the $M$-centered FSs possess electrons, 
but the other ones contain holes.
FSs of both systems are similar, but some considerable distinctions appear 
at the $M$-point and in the $\Gamma$-centered dumbbell-like pockets (see text).
The $M$-centered FSs connect to the $\Gamma$-centered large FS in \lnbn~ [see (a)],
while in \lpbn~ the $X$-centered FSs connect to each other [see (d)].
Fermi velocities are colored dark (blue) for the lowest value and lighter (red) for the highest.
}
\label{fs}
\end{figure}

\subsection{Fermi surface and Fermi velocity}
The Fermi surfaces (FSs) of \lnbn, displayed in Fig. \ref{fs}(a) and \ref{fs}(b),
consist of three $\Gamma$-centered electron surfaces, two $M$-centered electron surfaces, 
and a $X$-centered hole surface. 
In the $\Gamma$-centered FSs, there are cylindrical, dumbbell-like, and ellipsoidal pocket
FSs. Here, a tiny $\Gamma$-centered ellipsoid is invisible. 
The cylindrical pocket FS has a mixture of
La $5d_{x^2-y^2}$ and Ni $d_{z^2}$ orbitals, while the dumbbell-like one shows
La $5d_{x^2-y^2}$ and Ni $d_{xy}$ characters with small B and N $2p$ orbitals. 
The sandglass-like FS at the $M$ point shows substantial dispersion 
along the $k_z$ direction 
and has mainly Ni $3d_{xy}$ and $3d_{z^2}$ characters.
The $X$-centered electron FSs have mainly B $2p$ and N $2p$ characters.

Fig. \ref{fs}(c) and \ref{fs}(d) show FSs of \lpbn, which are similar to those of \lnbn. 
However, at the $M$-centered sandglass-like FS, which is much more dispersive along the $k_z$
direction, the character of Pt $5d$ orbitals mostly disappears.
The $\Gamma$-centered cylindrical FS, having mostly Pt $5d_{z^2}$ and La $5d_{x^2-y^2}$ characters
with small N $2p$ character, becomes more two-dimensional.
A most remarkable distinction appears in the $\Gamma$-centered dumbbell FS.
The dumbbell substantially reduces to a fishing-bob-like shape, which has mostly Pt $5d_{xy}$ character.
The small piece at the $M$ point disappears, but a small $Z$-centered starfish-like 
hole pocket appears.
Remarkably, although some parts of the $\Gamma$-centered FSs are parallel 
to another part on the $k_z=0$ plane in both systems, 
the strong three-dimensionality of the FSs may substantially reduce the strength of nesting, 
implying no spin or charge-density-wave instability.

To understand the transport properties, we have calculated the FS velocities,
as visualized in color in Fig. \ref{fs}.
In the dumbbell and bob FSs, in particular, the FS velocities are low  around the $\Gamma$ point.
The cylinders also have low FS velocity along the $\Gamma$-$M$ direction.
The root-mean-square obtained FS velocities $\langle v_{F,x}\rangle$ = 3.01 and 
$\langle v_{F,z}\rangle$ = 2.47 (in units of $10^7$ cm/s) for \lnbn.
Compared with the ratio $\langle v_{F,x}\rangle$:$\langle v_{F,z}\rangle$ of 2:1
in La$_3$Ni$_2$B$_2$N$_3$,\cite{wep95} these values are much less anisotropic.
The plasma energies are calculated from 
${\hbar\Omega_{p,ii}}=[\frac{4\pi e^2 \hbar^2 N(E_F)}{V_c} \langle v_{F,i}v_{F,i}\rangle]^{\frac{1}{2}}$,
where $V_c$ is the volume of the unit cell.
The plasma energies obtained are $\hbar\Omega_{p,xx}$ = 4.80 eV and $\hbar\Omega_{p,zz}$ = 3.95 eV.
\lpbn~ is more isotropic, with $\langle v_{F,x}\rangle$ = 3.50 and $\langle v_{F,z}\rangle$ = 3.80 
(in units of $10^7$ cm/s). The corresponding plasma energies are 
$\hbar\Omega_{p,xx}$ = 4.43 eV and $\hbar\Omega_{p,zz}$ = 4.80 eV.
As observed in the electronic structures, these indicate that
the transport is three-dimensional in both systems.

\begin{table*}[bt]
\caption{Optical phonon frequencies $\omega$ (cm$^{-1}$) at the $\Gamma$ point. 
Factor group analysis at the $\Gamma$ point produces 21 optical modes:
12 Raman active modes of 3$A_{1g}$ + $B_{1g}$ + 4$E_{g}$ and 9 infrared active modes
of 3$A_{2u}$ + 3$E_{u}$, where the $E_g$ and $E_{u}$ modes are 
two-fold degenerate.\cite{phonon1,phonon2}
Comparison shows remarkable variations in the Pt-related modes.
Interestingly, the last four modes have adjacent B and N atoms moving against each other. 
The atoms in parentheses show relatively small contributions to each mode.
The symbol $[110]^\ast$ indicates a vibration in an inclined direction 
toward one of the axes in the $ab$ plane. 
}
\begin{center}
\begin{tabular}{cccccccc}\hline\hline
   & \multicolumn{3}{c}{LaNiBN} &~&\multicolumn{3}{c}{LaPtBN}\\\cline{2-4} \cline{6-8}
 Symmetry & $\omega$ & Involved atoms & Vibration &~& $\omega$ & Involved atoms & Vibration\\\hline
$E_{g}$  & 112  & La, (B, N) & [110] &~& 110 & La, (B, N) & [100], [010]\\
$E_{u}$ & 133  & Ni, (La, B) & [110]&~& 95 & La, B, (N) & [110]$^\ast$\\
$A_{1g}$ &134  & La, (N)  & [001] &~& 149 & La, (B) & [001]\\
$A_{2u}$ &166  & Ni, (La, B, N) & [001]&~& 141 & La, Pt & [001]\\
$B_{1g}$ &194  & Ni          & [001] &~& 50 & Pt & [001]\\
$E_{g}$ &213   & Ni, (B, N)   & [110] &~& 204 & Pt, B, N & [110]$^\ast$\\
$E_{g}$ & 303  & N, (B)  & [110] &~& 284 & N, (B) & [100], [010]\\
$E_{u}$ & 311  & N, (B) & [110] &~& 283 & N, (B) & [110]$^\ast$\\
$A_{1g}$ & 339 & B, N  & [001] &~& 411 & B, N & [001]\\
$A_{2u}$ & 366 & B, N, (Ni) & [001] &~& 406 & B, N & [001]\\
$E_{u}$ & 398 & B, (N) & [110] &~& 436 & B, (N) & [110]$^\ast$ \\
$E_{g}$ & 448 & B, (N) & [110] &~& 465 & B, (N) & [110]$^\ast$\\
$A_{2u}$ & 938 & B, N & [001] &~& 992 & B, N & [001]\\
$A_{1g}$ & 948 & B, N & [001] &~& 1037 & B, N & [001]\\\hline 
\end{tabular}
\end{center}
\label{table1}
\end{table*}

\begin{figure}[tbp]
{\resizebox{8cm}{6cm}{\includegraphics{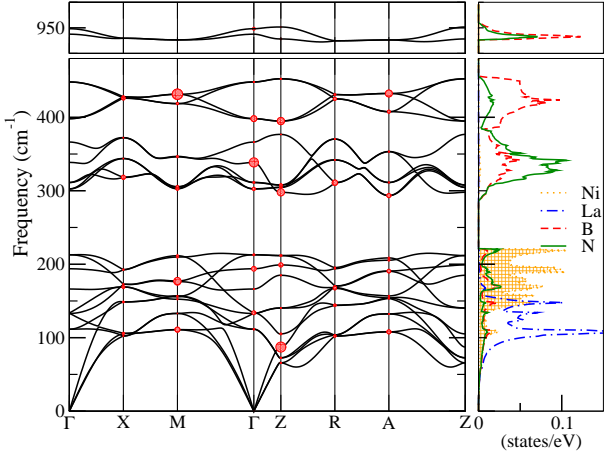}}}
\caption{(Color online) Phonon dispersion and atom-projected phonon DOSs
of \lnbn~ in the optimized structure.
In the phonon spectrum, the size of the red circles is proportional to the partial-mode
EPC strength of each phonon mode $\lambda_{\nu,\vec{q}}$,
almost uniformly distributed throughout most phonon modes, except for the highest modes.
}
\label{ph_ni}
\end{figure}

\begin{figure}[tbp]
{\resizebox{8cm}{6cm}{\includegraphics{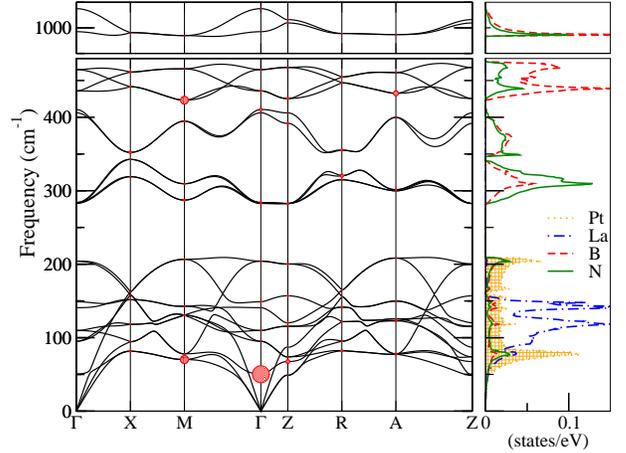}}}
\caption{(Color online) Phonon dispersion and atom-projected phonon DOSs of \lpbn~ 
in the optimized structure. 
In the phonon spectrum, the size of the red circles representing $\lambda_{\nu,\vec{q}}$
is reduced to half of that of \lnbn~ given in Fig. \ref{ph_ni} for better visualization.
Here, a few modes have relatively large $\lambda_{\nu,\vec{q}}$,
though most of all modes involved in EPC (for details, see text).
 }
\label{ph_pt}
\end{figure}

\begin{figure}[tbp]
{\resizebox{8cm}{6cm}{\includegraphics{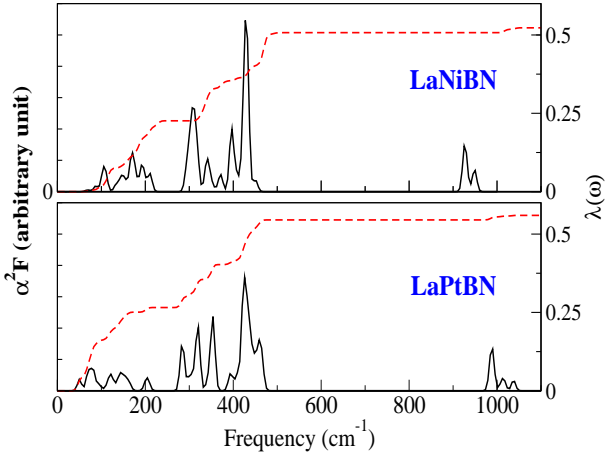}}}
\caption{(Color online) Eliashberg functions (left axis) and frequency-dependent EPC strengths
$\lambda(\omega)$ (right axis) of \lnbn~ and \lpbn~ in the optimized structure.
The root-mean-square phonon frequencies $\sqrt{\langle \omega^2\rangle}$ are 
184 and 162 K for \lnbn~and \lpbn, respectively, 
consistent with slightly stronger EPC in \lpbn.
}
\label{alpha}
\end{figure}

\section{phonon dispersion and electron-phonon coupling}
The phonon dispersion curve and the corresponding atom-projected phonon DOSs for \lnbn~
are displayed in Fig. \ref{ph_ni}.
The zone-centered phonon modes for both systems are analyzed in Table \ref{table1}.
These results are consistent with the frozen phonon calculations of La$_3$Ni$_2$B$_2$N, 
carried out by Singh and Pickett, who obtained the $A_{1g}$ Raman modes at 106, 323, and 896 
(in units of cm$^{-1}$).\cite{wep95}  
Low-energy phonon modes below 220 cm$^{-1}$ come from a mixture 
of vibrations of all atoms, but the Ni character appears mainly 
in the range of 140 to 220 cm$^{-1}$.
The higher phonon modes originate mostly from the vibrations of B and N ions.
The most unusual feature is the isolated modes at about 950 cm$^{-1}$
due to only B and N ions beating against each other, indicating
strong B-N bonding along the $c$ axis (see Table \ref{table1}).
This value is comparable to the highest B-C bonding stretching mode of $\sim$1200 cm$^{-1}$ 
in LiBC,\cite{an} though these beating modes do not contribute much 
to the superconductivity in \lnbn.

Fig. \ref{ph_pt} shows the phonon dispersion curve and the atom-projected phonon DOSs
of \lpbn, which are very similar to those of \lnbn.
Remarkably, the main peak of the Pt ion in the phonon DOS shifts down, 
since a Pt ion is heavier than a Ni ion.
As given in Table \ref{table1}, at the zone center the Pt $B_{1g}$ mode has a frequency 
of 50 cm$^{-1}$, which is a quarter of that of the Ni $B_{1g}$ mode but is
very similar to the frequency of the low-lying in-plane vibrational mode of the Pt ion  
in SrPt$_3$P.\cite{srpt3p,boeri}
This mode leads to slightly larger EPC strength in \lpbn~ (see below).
Additionally, compared with \lnbn, the frequency of the highest mode is enhanced 
by about 10\%.

Fig. \ref{alpha} shows the electron-phonon spectral function (Eliashberg function)
$\alpha^2F(\omega)$ and the EPC strength $\lambda$ together for both systems, 
which are given by\cite{Mcmillan}
\begin{eqnarray}
 \alpha^2F(\omega)=\frac{1}{2N} \sum_{\nu,\vec{q}} \omega_{\nu,\vec{q}}
 \lambda_{\nu,\vec{q}} \delta(\omega -\omega_{\nu,\vec{q}}), \\
 \lambda = 2\int_0^{\infty} d\omega \frac{\alpha^2F(\omega)}{\omega} 
        = \frac{N(E_F)\langle I^2\rangle}{M\langle \omega^2\rangle},                       
\end{eqnarray}
where $\langle I^2\rangle$ is the mean-square electron-phonon matrix element averaged
over FS and $M$ is the ion mass.  
$\lambda$ can also be obtained as an average of $\lambda_{\nu,\vec{q}}$
over $N$ $q$ mesh and all the phonon modes, 
$\lambda=\frac{1}{N} \sum_{\nu,\vec{q}} \lambda_{\nu,\vec{q}}$.
$\lambda_{\nu,\vec{q}}$ are visualized as red circles in Fig. \ref{ph_ni} and \ref{ph_pt},
indicating the mode EPC distributes throughout almost all phonon modes, 
except for the highest modes, 
which is consistent with the corresponding Eliashberg functions. 
(Note that the size of the circles in Fig. \ref{ph_ni} is 
two times larger than in Fig. \ref{ph_pt} for better visualization.)
Our approach is different from previous studies, which focused on the B $A_{1g}$ mode
for $n=1$ borocarbides and $n=3$ boronitrides.\cite{wep95,matt} 
Our result is in good agreement with the recent experiment using a time-of-flight technique.\cite{weber}
In \lpbn, a few modes showing some large $\lambda_{\nu,\vec{q}}$ appear 
in the $B_{1g}$ mode at the $\Gamma$ point, 
in the two fold degenerate modes around 70 and 400 $cm^{-1}$ at the $M$ point, 
and around 400 $cm^{-1}$ at the $A$ point,
but each of these modes equally contributes only 8\% to the $\lambda$.
The two lower energy modes are involved in the Pt vibrations along the $c$ axis, 
but the other modes are due to the B-N out-of-phase vibrations in the $ab$ plane. 
In spite of these distinctions, the EPC strengths of the two systems are very close, 
with $\lambda$ = 0.52 and 0.56 for \lnbn~and \lpbn, respectively. 
Although the main peaks in $\alpha^2F(\omega)$ that are mostly attributed to the vibrations of B and N atoms 
appear in the range of 250 to 500 $cm^{-1}$, their contributions to $\lambda$ are only about 50\%.
The other half is contributed by low-frequency phonon modes below 250 cm$^{-1}$ due to
the factor of $\frac{1}{\omega}$ in $\lambda$, as shown in Eq. (2).
To calculate $T_c$, we used the Allen-Dynes equation,\cite{allen} neglecting
insignificant strong-coupling corrections in this case, 
\begin{eqnarray}
 T_c = \frac{\omega_{log}}{1.2} {\rm exp}[-\frac{1.04(1+\lambda)}{\lambda-\mu^\ast-0.62\lambda\mu^\ast}],
\end{eqnarray}
where $\mu^\ast$ is an effective Coulomb repulsion, which is uncertain for these systems
but is typically in the range of 0.1 -- 0.2. 
The logarithmically averaged characteristic phonon frequencies $\omega_{log}$ are
376 and 294 K for \lnbn~ and \lpbn, respectively.
Thus, $T_c \approx$ 3.9 K with $\mu^\ast$ = 0.12 for \lnbn~ and 5.4 K with $\mu^\ast$ = 0.1 for \lpbn,
which are very close to the observed values.
Considering the similar $N(E_F)$ and larger bandwidth of the Pt $5d$ orbital, 
these choices of $\mu^\ast$ are reasonable.
As a result, the conventional EPC model effectively explains the superconductivity of both systems.

Furthermore, by changing the mass $M$ of the B atom a little,
we obtained an isotope shift $\alpha_B=-d(log T_c)/d(log M)$ of $-0.15$ for \lnbn~ 
and 0.13 for \lpbn, which are similar to the experimentally measured values 
of 0.1 -- 0.25 in a few borocarbides.\cite{iso1,iso2,iso3}
These values are much smaller than the BCS value of 0.5.
This fact is consistent with the fact that the superconductivity is not purely mediated 
by B-related vibrational modes.
The negative isotope shift of B in \lnbn~ seems to be unusual 
but was also suggested through the Raman scattering measurements of 
the nickel borocarbides.\cite{canfield}

\begin{figure}[tbp]
{\resizebox{8cm}{6cm}{\includegraphics{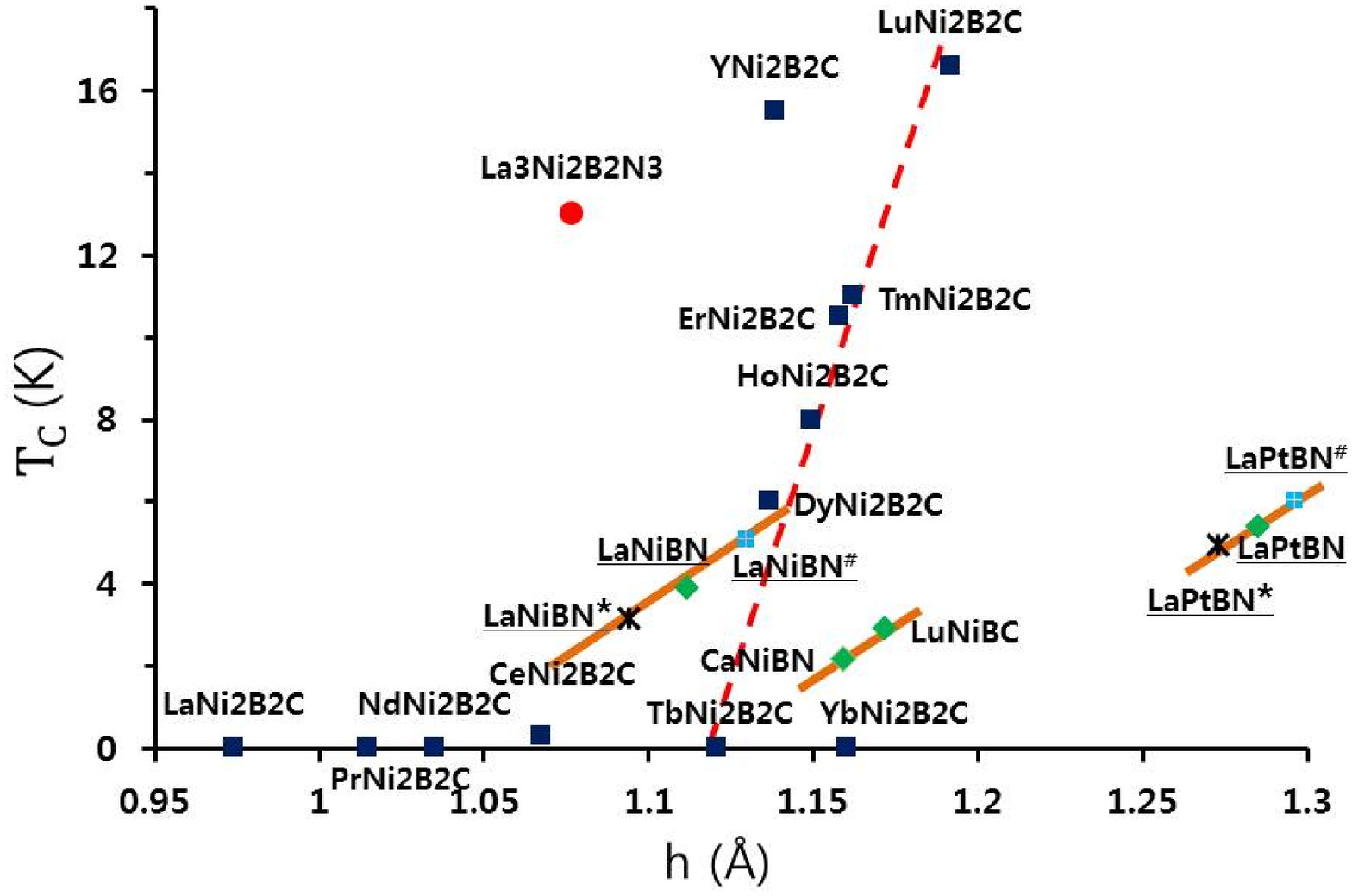}}}
\caption{(Color online) The relation between 
the superconducting critical temperature $T_c$ and
the B height $h$ in the ${\cal M}$-B layer, as displayed in Fig. \ref{str}, 
in borocarbides and boronitrides.
Here, $n=1$ and $n=2$ systems are denoted by symbols of filled squares 
and filled diamonds, respectively. 
The superconducting $n=1$ borocarbides are connected by the (red) dashed line.
The solid lines connect two mutually related compounds in $n=2$ systems.
The data for \lnbn$^\ast$ (\lpbn$^\ast$) and \lnbn$^\#$ (\lpbn$^\#$)
are obtained from our calculations for 3\% compressed and 3\% enlarged volumes,
respectively.
The data are from Ref. [\onlinecite{wep95}] for La$_3$Ni$_2$B$_2$N$_3$, 
Ref. [\onlinecite{ynbc_n1}] for YNi$_2$B$_2$C, Ref. [\onlinecite{lanbc_n1}]
for LaNi$_2$B$_2$C and LuNiBC, Ref.[\onlinecite{lunbc_n1}] for LuNi$_2$B$_2$C,
and Ref. [\onlinecite{rnbc_n1}] for other $n=1$ borocarbides.
}
\label{geom}
\end{figure}

\section{Discussion and Summary}
As a starting point to anticipate the effects of pressure on $T_c$,
we reduced the volume by 3\%.
(During this reduction, the internal parameters were optimized with the ratio of $a/c$,
fixed to that of the uncompressed structure.)
Although the electronic structures in both systems remain nearly unchanged 
in almost all regimes around $E_F$, noticeable variations occur 
in the top of the valence (occupied) bands at the $\Gamma$ and $Z$ points.
In \lnbn, at the $\Gamma$ point one of the top valence bands becomes unoccupied,
so that the dumbbell FS is nearly separated into two ellipsoids.
The phonon frequencies shift up in the phonon spectrum, indicating
the lattice becomes stiffer, as confirmed by enhancing $\omega_{log}$.
This leads to decreasing $\lambda$ and finally reducing $T_c$ by 18\%.
Changes in \lpbn~ are very similar to those in \lnbn.
The flat band along the $\Gamma$-$Z$ line is shifted above $E_F$,
so the bob-shaped FS disappears.
As expected from the small size of the bob, reduction in $T_c$ of \lpbn~ 
is less significant, only 8\%.
This may imply less dramatic variation in the superconductivity 
under pressure in \lpbn.

A remarkable relation between $T_c$ and the geometric factor has been sometimes
observed, {\it e.g.}, Fe pnictides showing substantial dependence 
of $T_c$ on the As-Fe-As bond angle.\cite{clee}
Also in the $n=1$ ${\cal R}$Ni-based borocarbides, 
a linear dependence of $T_c$ on the ratio of 
${\cal R}$-C distance to lattice parameter $a$ was suggested, 
implying the EPC mediated superconductivity.\cite{BC1}
Interestingly, we found another interesting trend between $T_c$ and 
the B height $h$ in the ${\cal M}$-B layer, which shows a monotonic increase of $T_c$
with respect to $h$, as shown in Fig. \ref{geom}.
$T_c$ of $n=1$ borocarbides shows a very steep slope of $dT_c/dh$ = 2.27 K/pm,
but no superconductivity has been observed for $h$ below 1.12 \AA. 
One may expect similar behavior for $n=2$ or $n=3$ systems,
even though the existing data are limited.
Using the data for the experimental and $\pm$3\% varied volumes for \lnbn~ and \lpbn,
we obtained a less rapid increment of $T_c$ with a nearly identical slope of 
$dT_c/dh$ = 0.52 ($\pm$0.02) K/pm.
As addressed above in \lnbn~ and \lpbn, 
different topology and size of one of the $\Gamma$-centered FSs
among the compounds along each line in Fig. \ref{geom} may lead to this change in $T_c$.
These trends suggest that $T_c$ of \lnbn~ and \lpbn~ could be enhanced 
by replacing La by a bigger ${\cal R}$ atom.  

In summary, we have investigated the electronic structures and phonon properties 
of two superconductors, \lnbn~ and \lpbn, which have 
structures similar to those of the superconducting Fe pnictides, 
through first-principles approaches.
Our results indicate insignificant effects of SOC, 
and there is no evidence of magnetic instability in either system.
These systems show considerable three-dimensionality 
with both metallic La-N and ${\cal M}$-B layers, 
but relatively low DOS at $E_F$. Both systems have very similar phonon spectra.
One remarkable distinction is a significant reduction in the {$\cal M$} $B_{1g}$ mode in \lpbn, 
resulting in the enhancement of $\lambda$ by about 8\% compared to the $\lambda$ = 0.52 value of \lnbn.
These EPC strengths lead to $T_c$, consistent with the experimental values. 
Thus, our findings suggest that the superconductivity is mediated 
by conventional electron-phonon coupling.

The comparison between our calculated $T_c$ and the experimentally observed value in \lpbn~
shows some room, which may be filled by additional interactions like spin fluctuation.
However, the rather small contribution of the Pt ion to the DOS near $E_F$ 
indicates that such a magnetic interaction is improbable in \lpbn.
Further experiments would be desirable to clarify this feature in \lpbn.

\section{Acknowledgments}
We acknowledge W. E. Pickett and D. J. Scalapino for fruitful discussions 
and K. Keopernick for providing a program to calculate the Fermi surface and Fermi velocity. 
This research was supported by the Basic Science Research Program through
the NRF of Korea under Grant No. 2012-0002245.
C.J.K. and B.I.M. were supported by the NRF under Grant No. 2009-0079947.

\end{document}